\documentclass{aa}
\usepackage{txfonts}
\usepackage{graphicx}
\usepackage{psfig,longtable,lscape}
\usepackage{epsfig}

\begin{document}

\title{A ionized reflecting skin above the accretion disk of GX 349+2}
  

\author{R. Iaria\inst{1}, A. D'A\'\i\inst{1}, T. Di Salvo\inst{1}, 
N. R. Robba\inst{1}, A. Riggio\inst{2}, A. Papitto\inst{3}$^,$\inst{4}, L. Burderi\inst{2}}

\offprints{R. Iaria, \email{iaria@fisica.unipa.it}}

\institute{Dipartimento di Scienze Fisiche ed Astronomiche,
Universit\`a di Palermo, via Archirafi 36 - 90123 Palermo, Italy
  \and 
Dipertimento di Fisica, Universit\`a degli Studi di Cagliari, SP Monserrato-Sestu, KM 0.7, Monserrato, 09042 Italy
  \and 
Dipartimento di Fisica, Universit\`a  degli Studi di Roma 'Tor Vergata', via della Ricerca Scientifica 1,00133 Roma,Italy
  \and 
INAF Osservatorio Astronomico di Roma, via Frascati 33, Monteporzio Catone, 00040, Italy
} 

\date{}

\abstract
{The broad emission features in the Fe-K$\alpha$ region of X-ray
  binary spectra represent an invaluable probe to constrain the
  geometry and the physics of these systems.  Several Low Mass X-ray
  binary systems (LMXBs) containing a neutron star (NS) show broad
  emission features between 6 and 7 keV and most of them are now
  interpreted as reflection features from the inner part of an
  accretion disk in analogy to those observed in the spectra of X-ray
  binary systems containing a Black Hole candidate.}
{The NS LMXB  GX 349+2 was observed by the XMM-Newton  satellite which
  allows, thanks  to its high effective  area and good spectral resolution 
between 6 and  7 keV, a
  detailed spectroscopic study of the Fe-K$\alpha$ region.}
{ We study the XMM data in the 0.7-10 keV energy band. The continuum
  emission is modelled by a blackbody component plus a multicolored
  disk blackbody.  A very intense emission line at 1 keV, three broad
  emission features at 2.63, 3.32, 3.9 keV and a broader emission
  feature in the Fe-K$\alpha$ region are present in the residuals.
  The broad emission features above 2 keV can be equivalently well
  fitted with Gaussian profiles or relativistic smeared lines ({\tt
    diskline} in XSPEC). The Fe-K$\alpha$ feature is better fitted
  using a diskline component at 6.76 keV or two diskline components at
  6.7 and 6.97 keV, respectively}
{ The emission features are interpreted as resonant transitions of
  \ion{S}{xvi}, \ion{Ar}{xviii}, \ion{Ca}{xix}, and highly ionized iron. 
 Modelling the line profiles with relativistic
  smeared lines, we find that the reflecting plasma is located at less
than 40  km from the NS, a value compatible with the inner
  radius of the accretion disk inferred from the multicolored disk
  blackbody component ($24 \pm 7$ km).  The inclination angle of
  GX 349+2 is between 40$^\circ$ and 47$^\circ$, the emissivity index
  of the primary emission is between -2.4 and -2, and the reflecting
  plasma extends up to (2-8) $\times 10^8$ cm.}
{We compare our results with the twin source Sco X-1 and with the
  other NS LMXBs showing  broad relativistic lines in their spectra.
  We conclude that the blackbody component in the spectrum is the
  primary emission that hits the inner accretion disk producing the
  emission lines broadened by relativistic and Doppler effects
  dominant around the neutron star.}
\authorrunning {R.\ Iaria et al.}
\titlerunning {A ionized reflecting skin in GX 349+2}

\maketitle

\section{Introduction}

Low-mass X-ray binaries (LMXBs) consist of a low-mass star ($M<1
M_{\odot}$) and a neutron star (NS), which generally has a relatively weak
magnetic field ($B<10^{10}$G). In these systems, the X-ray source is
powered by accretion of mass overflowing the Roche lobe of the
companion star and forming an accretion disk around the neutron star.
LMXBs containing a neutron star (NS LMXBs) are generally divided into
Z and Atoll sources, according to the path they describe in an X-ray
Color-Color Diagram (CD) or hardness-intensity diagram
\citep{HasVdKlis89} assembled by using the source count rate and
colors calculated over a typical (usually 2-20 keV) X-ray energy
range.  Atoll sources are usually characterized by relatively low
luminosities (0.01-0.2 L$_{Edd}$ ) and often show transient behavior,
while the Z sources in the Galaxy are among the most
luminous LMXBs,  persistently accreting close to the Eddington limit
(L$_{Edd}$) for a 1.4 M$_{\odot}$ NS. The position of an
individual source in the CD, which determines most of the observed
spectral and temporal properties of the source, is thought to be an
indicator of the instantaneous mass accretion rate \citep[e.g.][for a
review]{Has90,VdKlis95}.  It has been suggested that the mass
accretion rate (but not necessarily the X-ray luminosity) of
individual sources increases along the track from the top left to the
bottom right, i.e.  from the islands to the banana branch in atoll
sources and from the horizontal branch (hereafter HB) to the normal
branch (NB) and to the flaring branch (FB) in Z sources.

   GX 349+2, also known as Sco X-2, was called an odd-ball among the Z
   sources \citep{KuulVdk}. Similar  to the case of Sco  X-1, GX 349+2
   shows  a short  and  underdeveloped  HB (if  at  all).  The  source
   variability  in  the  frequency  range  below  100  Hz  is  closely
   correlated with the source position on  the X-ray CD, as in other Z
   sources.  Quasi periodic oscillations  at kHz frequencies (kHz QPO)
   were detected in the  NB of its Z-track \citep*{Zhang98}.  However,
   GX 349+2, which sometimes shows broad noise components changing not
   only with  the position  in the Z,  but also  as a function  of the
   position in  the hardness-intensity diagram,  differs somewhat from
   the other Z sources and  shows similarities to the behavior seen in
   bright   atoll   sources,   such    as   GX   13+1   and   GX   3+1
   \citep{KuulVdk,oneill01,oneill02}.  

   \cite{DiSalvo01}, using BeppoSAX data, showed that the source
   energy spectrum below 30 keV could be well fit by a blackbody (with
   a temperature of 0.5-0.6 keV) and a Comptonized component (with
   seed-photon temperature of 1 keV and electron temperature of 2.7
   keV).  Three discrete features were observed in the spectrum: an
   emission line at 1.2 keV, probably associated to L-shell
   \ion{Fe}{xxiv} or Ly-$\alpha$ \ion{Ne}{x}, an emission line at 6.7
   keV (\ion{Fe}{xxv}) and an absorption edge at 8.5 keV, both
   corresponding to emission from the K-shell of highly-ionized iron
   (\ion{Fe}{xxv}).  \cite{Iaria04}, analysing a long BeppoSAX
   observation, found similar parameters of the the continuum
   components below 30 keV, detected the presence of emission lines
   associated to \ion{Fe}{xxiv} and \ion{Fe}{xxv}, and detected an
   emission line at 2.65 keV associated to \ion{S}{xvi}.  The long
   BeppoSAX observation allowed to study changes of the parameters of
   the \ion{Fe}{xxv} emission line along the position of the source in
   the CD from the NB/FB apex to the FB inferring that the equivalent
   width of the line decrease from 77 eV to 18 eV going from the NB/FB
   apex to the FB.  The width of the \ion{Fe}{xxv} emission line,
   modelled by a Gaussian line, was between 250 and 300 eV.
\begin{figure}
\includegraphics[height=8.cm,angle=0]{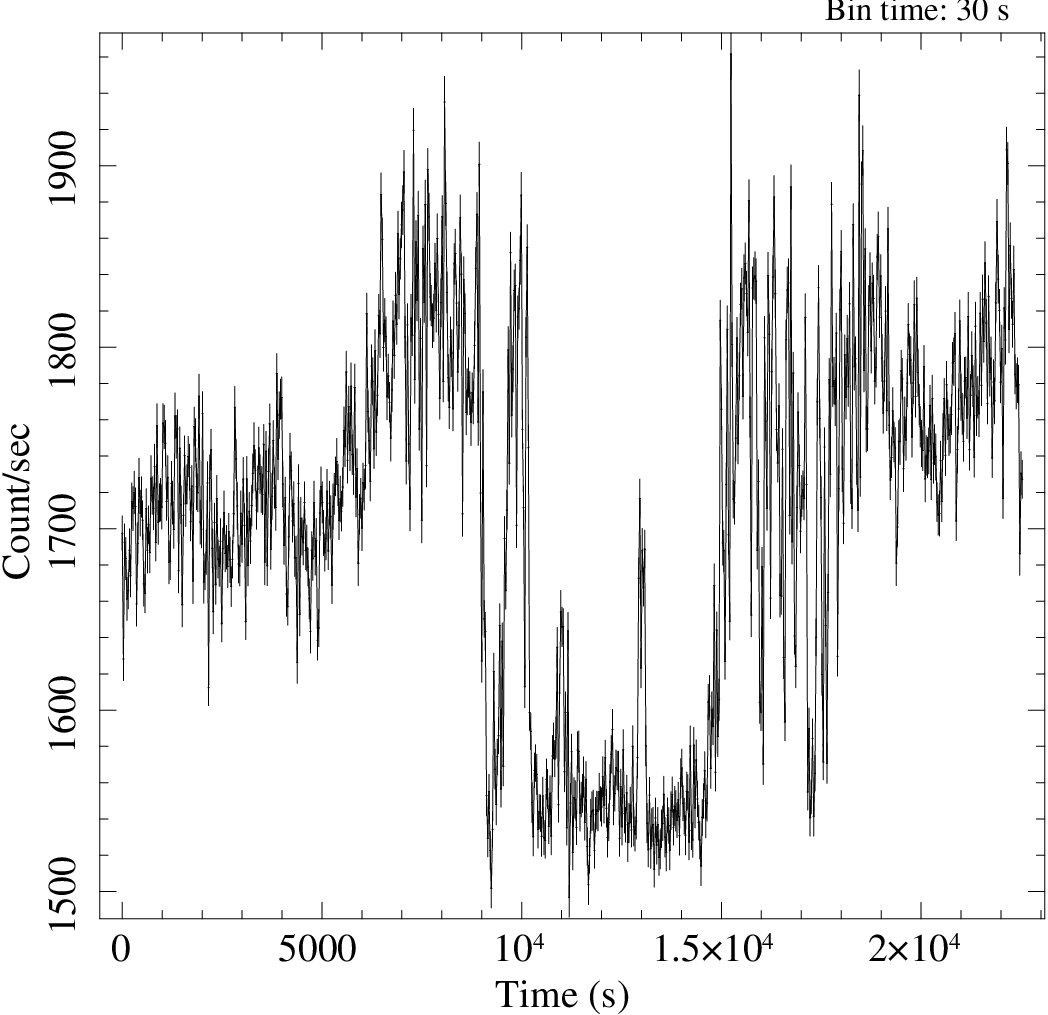}
\caption{Epic-pn lightcurve  of GX 349+2  in the 
energy  band 1-10 keV. The bin time is 30 s.}
\label{pino}
\end{figure}
   
Recently \cite{Cackett08_349}, analysing  two Suzaku observations of GX
349+2, modelled the \ion{Fe}{xxv} emission line with a relativistic
smeared line (diskline in  XSPEC).  The authors constrained the energy
of the line to be between 6.4 and  6.97 keV finding that
the  line  energy  was  $6.97_{-0.02}$  keV,  associated to \ion{Fe}{xxvi} 
and  not  compatible  with  a
\ion{Fe}{xxv} iron  line as previously obtained  by \cite{Iaria04} and
\cite{DiSalvo01}.  The emissivity index,  $\beta$, was $-4.1 \pm 0.3$,
the inner radius in gravitational radii, $R_g$, was $8.0 \pm 0.4$, the
inclination angle of the source was $23 \pm 1$ degrees, the equivalent
width of the line was $76 \pm 6$ eV.

In this paper  we analyse an XMM-Newton observation of GX 349+2
finding that the prominent relativistic line is well constrained
around 6.7 keV and identified as the resonant transition of
\ion{Fe}{xxv}.  In the following we discuss our results and compare
them with the recent literature.

 \section{Observation}
 
\begin{figure}
\includegraphics[height=8.cm,angle=0]{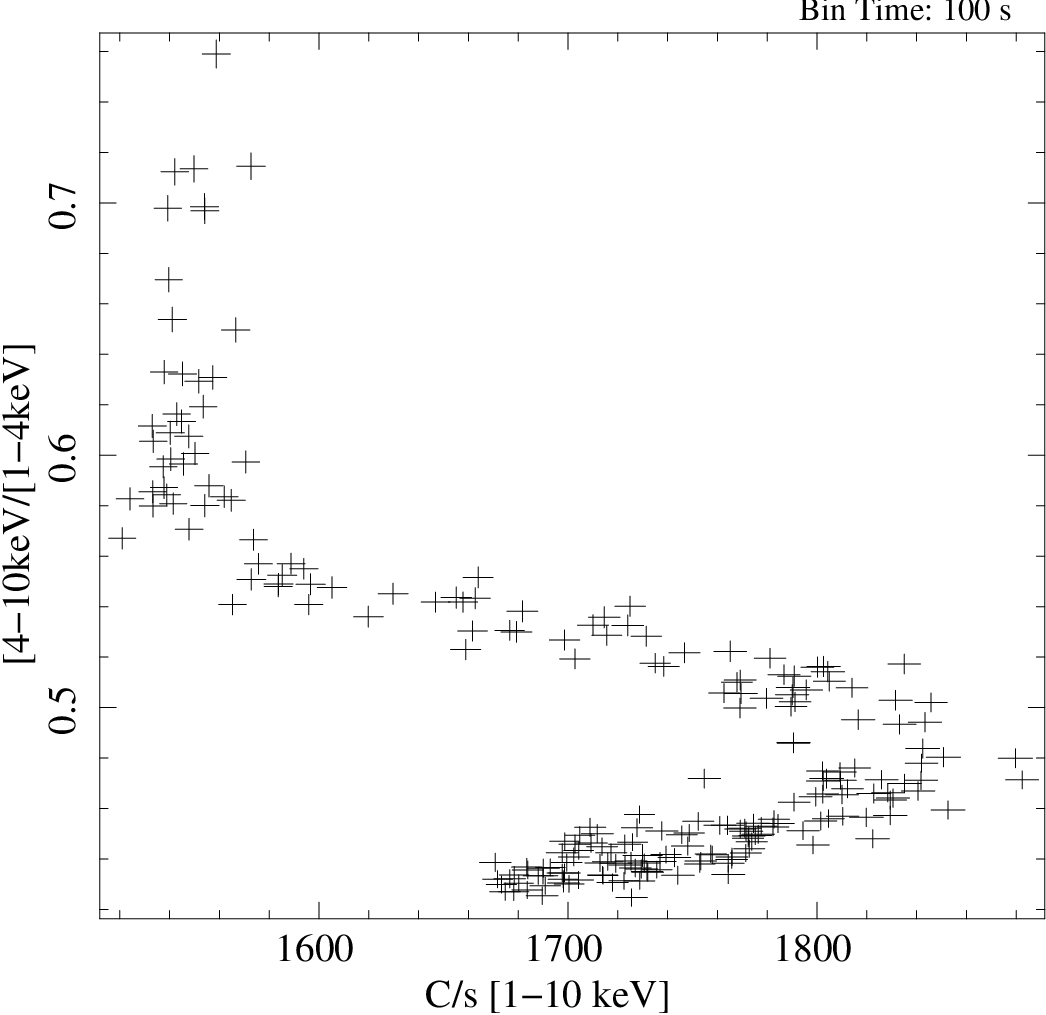}
\caption{ Color-Intensity diagram of GX 349+2. 
The [4-10keV]/[1-4keV] hardness ratio increases for a count rate 
lower than 1580 c/s. The bin time is 100 s.}
\label{pino2}
\end{figure}
The XMM-Newton Observatory \citep{jansen01} includes three 1500 cm$^2$
X-ray telescopes each with an European Photon Imaging Camera (EPIC,
0.1--15 keV) at the focus.  Two of the EPIC imaging spectrometers use
MOS CCDs \citep{turner01} and one uses pn CCDs \citep{struder01}.
Reflection Grating Spectrometers \citep[RGS, 0.35--2.5
keV,][]{denherder01} are located behind two of the telescopes.  The
region of sky containing GX 349+2 was observed by XMM-Newton between
2008 March 19 16:42:41 UT to March 19 22:58:55 UT (OBSid 0506110101)
for a duration of 22.5 ks. Since GX 349+2 is one of the brightest
X-ray sources, in order to minimize the effects of telemetry
saturation and pile-up the MOS1 and MOS2 instruments were switched off
and the EPIC-pn camera was operated in Timing mode with medium filter
during the observation. In this mode only one central CCD is read out
with a time resolution of 0.03 ms.  This provides a one dimensional
(4\arcmin.4 wide) image of the source with the second spatial
dimension being replaced by timing information.  The faster CCD
readout results in a much higher count rate capability of 1500 cts/s
before charge pile-up become a serious problem for point-like
sources. The EPIC-pn telemetry limit is approximatively 450 c/s for
the timing mode\footnote{see Tab. 3 in XMM-Newton Users Handbook
  published on 15 July 2008}. If the rate is higher, then the counting
mode is triggered and part of the science data are lost.  This is the
case of the observation presented in this work; saturation occurs for
a 55\% of the observing time,
 giving an exposure time of 22.5 ks for RGS1 and RGS2, respectively, and
 10 ks for EPIC-pn.

 We plot the Epic-pn lightcurve in the 1-10 keV energy range in Fig.
 \ref{pino} which clearly shows that  the count rate is largely variable.
To understand whether the variability of the source
 intensity is energy dependent  we extract the lightcurves in
 the 1-4 keV and 4-10 keV energy range, respectively, and plot the
 hardness ratio [4-10 keV]/[1-4keV] versus the count rate in the 1-10
 keV energy range in Fig. \ref{pino2}.
 We find that the spectrum becomes harder for a count rate below 1580
 c/s corresponding to the time interval between 10$^4$ and $1.5 \times
 10^4$ s from the start of the observation.
\begin{figure}
\includegraphics[height=8.5cm,angle=0]{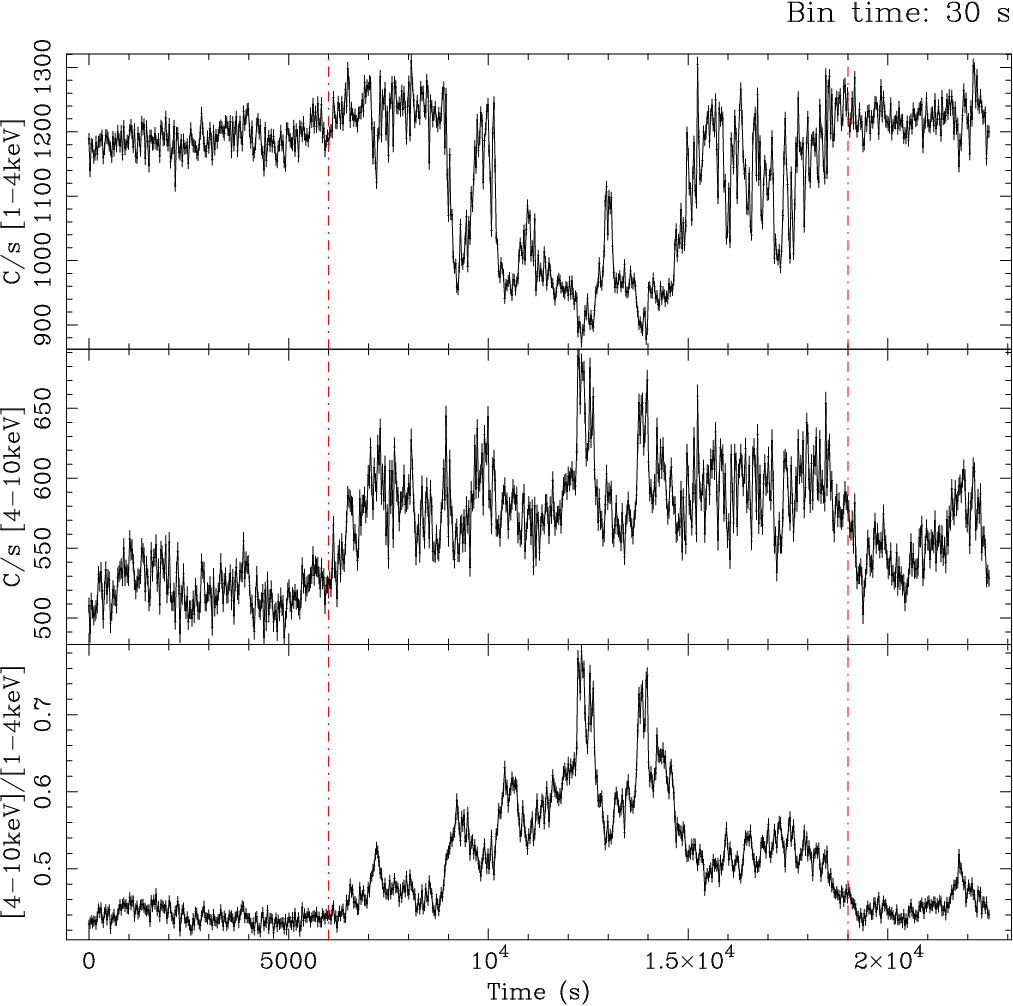}
\caption[]{Upper Panel: 1-4 keV Epic-pn lightcurve. Middle Panel:
 4-10 keV Epic-pn lightcurve. Bottom panel: [4-10keV]/[1-4keV] hardness ratio.
The bin time is 30 s for each lightcurve. The red dot-dashed vertical lines 
indicate the three selected time intervals.  }
\label{hardness}
\end{figure}  
Plotting the 1-4 keV, 4-10 keV lightcurve and the [4-10 keV]/[1-4keV]
hardness ratio versus time (see Fig. \ref{hardness}), it seems that
there is no correlation between the 1-4 keV and 4-10 keV count rate.
The 1-4 keV count rate is quite constant at 1200 c/s between 0 and $9
\times 10^3$ s and between $1.5 \times 10^4$ and $2.2 \times 10^4$ s
from the start, and it decreases between $9 \times 10^3$ and $1.5
\times 10^4$ s, showing some flaring activity.  The 4-10 keV count rate
is quite constant at 530-550 c/s between 0 and $6 \times 10^3$ s and
between $1.9 \times 10^4$ and $2.25 \times 10^4$ s, while increases to
600 c/s between $6 \times 10^3$ and $1.9 \times 10^4$ s.

Since the count rate of the source is variable,  we divide the lightcurve into
 three time intervals as
indicated in Fig. \ref{hardness} by the red dot-dashed lines. We base
our choice on the 4-10 keV count rate that is 550 c/s in the first and
third time interval and it is around 600 c/s in the second time
interval. Our choice allows to highlight  possible changes in the
Fe-K$\alpha$ region of the spectrum.   In Tab. \ref{tableLOGObs} we
report the exposure times of the selected time intervals for RGS1, RGS2
and EPIC-pn instruments.

\begin{table}[ht]
  \caption{Time intervals selected in the EPIC-pn lightcurve}
\label{tableLOGObs}      
\begin{center}                                      
\begin{tabular}{l c c c}          
\hline\hline 
 &
Time interval&
RGS1&
EPIC-pn\\                       
 &
s - TSTART[s]&
ks &
ks \\
\hline

INT. 1 &
0-6000 &
6.0  &
2.6
\\
INT. 2 &
6000-19000& 
13.0& 
5.8 \\

INT. 3 &
$>$19000& 
3.5& 
1.6 \\
\hline                                             
\end{tabular}
\end{center}                                      

{\small \sc Note} \footnotesize---  The observation was divided into three
time intervals as shown in Fig. \ref{hardness}. In the first column we report 
the temporal  boundaries of each interval  with respect 
 to the start 
time of the observation. In the second and third column we 
report the exposure times, in kiloseconds, for RGS1, RGS2, and EPIC-pn during 
the three time intervals.        
 \end{table}

\subsection{Spectral analysis of the averaged spectrum}
We extract the X-ray data products of RGS and EPIC-pn camera using the
Science Analysis Software (SAS) version 8.0.0 extracting only single
and double events (patterns 0 to 4) from EPIC-pn data.  Initially we
extracted source EPIC-pn events from a $69.7\arcsec$ wide column (RAWX
between 29 and 45) centered on the source position (RAWX$=37$).
Background events were obtained from a box of the same width
with  RAWX between 2 and 18.  

 The 0.5-12 keV Epic-pn count rate extracted from the source region is
almost 1800 c/s, since the maximum Epic-pn count rate should be 800
c/s to avoid a deteriorated response due to photon
pile-up\footnote{see Tab. 3 in XMM-Newton Users Handbook published on
  15 July 2008}, we expect presence of pile-up. We checked the
pile-up fraction at different energies using the {\it epatplot} tool
in SAS
\begin{table}[ht]
  \caption{Percentage of photon pile-up at different energies}
\label{pileup}      
\begin{center}                                      
\begin{tabular}{l  c  c }          
\hline\hline 
Energy  &
Pile-up (\%) & 
Pile-up (\%)\\
    (keV)         &
 RAWX [29:45] & 
 RAWX [29:36] and RAWX[39:45]\\

\hline

5 &
$1.3 \pm 0.1$  &
$0.2 \pm 0.2$
\\
6 &
$2.0 \pm 0.1$& 
$0.6 \pm 0.2$ \\

7 &
$3.2 \pm 0.2$& 
$1.6 \pm 0.2$ \\

8 &
$5.1 \pm 0.3$& 
$1.9 \pm 0.4$ \\

9 &
$7.4 \pm 0.4$& 
$4.8 \pm 0.5$ \\

10 &
$9.8 \pm 0.6$& 
$6.5 \pm 0.8$ \\
\hline                                             
\end{tabular}
\end{center}                                      

{\small \sc Note} \footnotesize---  The associated errors are reported at 1 $\sigma$.
 In the first column we report the energies, in the second,  and third
 column
the estimated percentage of photon pile-up extracting the events from 
the whole source region and excluding the brightest columns 
at  RAWX$=37$ and  RAWX$=38$,  respectively.      
 \end{table}
\begin{table}[h]
  \caption{Averaged Continuum and Identified Broad Gaussian Lines below 6 keV.}
\label{table:model_gauss}      
\begin{center}
\begin{tabular}{l c c}          
\hline\hline                        
Parameters  &
Single Fe-K$\alpha$ &
\ion{Fe}{xxv}-\ion{Fe}{xxvi}  \\
  &
Em. Line &
Blending \\

\hline                        
&&\\

N$_H$  (10$^{22}$ cm$^{-2}$)&  
$0.735 ^{+0.006}_{-0.003}$&
$0.736 ^{+0.005}_{-0.004}$\\

&&\\

kT$_{BB}$ (keV)         &
$1.792 ^{+0.006}_{-0.019}$&
$1.787 ^{+0.015}_{-0.014}$\\

N$_{BB}$ (10$^{-1}$) &
$1.052 ^{+0.017}_{-0.009}$&
$1.057 ^{+0.012}_{-0.014}$\\
&&\\

kT$_{DISKBB}$ (keV)         &
$1.05^{+0.02}_{-0.03}$&
$1.04 \pm 0.03$\\

N$_{DISKBB}$  &
$210^{+10}_{-30} $&
$220 \pm 20 $\\
&&\\

E$_{1 keV}$ (keV) &
  $1.051^{+0.006}_{-0.015}$&
 $1.051^{+0.006}_{-0.015}$  \\

$\sigma_{1 keV}$ (eV) &
$90^{+11}_{-8}$   &
$85^{+12}_{-6}$  \\

I$_{1 keV}$ (10$^{-3}$ ph cm$^{-2}$ s$^{-1}$) & 
$25 \pm 3$   &
$24^{+4}_{-2}$   \\

Equiv. width (eV) & 
$22^{+4}_{-3}$  &
$21 \pm 2$  \\

significance ($\sigma$) &
14 &
18 \\

&&\\

E$_{S \; XVI}$ (keV) &
2.62 (fixed)  &
2.62 (fixed)  \\

$\sigma_{S \; XVI}$ (eV) &
140 (fixed)  &
140 (fixed)  \\

I$_{S \; XVI}$ (10$^{-3}$ ph cm$^{-2}$ s$^{-1}$) & 
$2.3^{+0.9}_{-0.5}$   &
$2.5^{+0.7}_{-0.6}$   \\

Equiv. width (eV) & 
$6 \pm 2$  &
$6 \pm 2$  \\

significance ($\sigma$) &
7 &
7 \\

&&\\

E$_{Ar \; XVIII}$ (keV) &
$3.33^{+0.03}_{-0.04}$  &
$3.32 ^{+0.03}_{-0.04}$  \\

$\sigma_{Ar \; XVIII}$ (eV) &
$190^{+70}_{-40}$  &
$200^{+60}_{-50}$  \\

I$_{Ar \; XVIII}$ (10$^{-3}$ ph cm$^{-2}$ s$^{-1}$) & 
$3.7^{+1.7}_{-0.4}$   &
$4.0^{+1.4}_{-1.1}$   \\

Equiv. width (eV) & 
$12 \pm 5$  &
$14^{+4}_{-5}$  \\

significance ($\sigma$) &
15 &
6 \\

&&\\

E$_{Ca \; XIX}$ (keV) &
$3.93^{+0.03}_{-0.02}$ &
$3.93 \pm 0.02$ \\

$\sigma_{Ca \; XIX}$ (eV) &
$100^{+50}_{-30}$  &
$110^{+50}_{-20}$  \\

I$_{Ca \; XIX}$ (10$^{-3}$ ph cm$^{-2}$ s$^{-1}$) & 
$2.2^{+0.7}_{-0.4}$   &
$2.3^{+0.6}_{-0.4}$   \\

Equiv. width (eV) & 
$9 \pm 3$  &
$9^{+3}_{-2}$  \\
significance ($\sigma$) &
9 &
9 \\

&&\\

$\chi_{red}^2(d.o.f.)$ &
$1.51(269)$  &
$1.51(268)$  \\

\hline                                             
\end{tabular}
\end{center}

{\small \sc Note} \footnotesize---  Best-fit values of the parameters obtained
for the averaged spectrum fitting the
emission features with Gaussian profiles.   Uncertainties are  at the  90\% 
confidence
level for a single parameter.   
The parameters are defined as in XSPEC. The continuum emission is fitted with 
a {\tt blackbody} plus a multicolored disk blackbody, {\tt diskbb}. 
The four broad emission features are modelled adopting {\tt Gaussian} 
components.  All components are subject to interstellar absorption. 
\end{table}
\begin{figure}
\includegraphics[height=8.cm,angle=0]{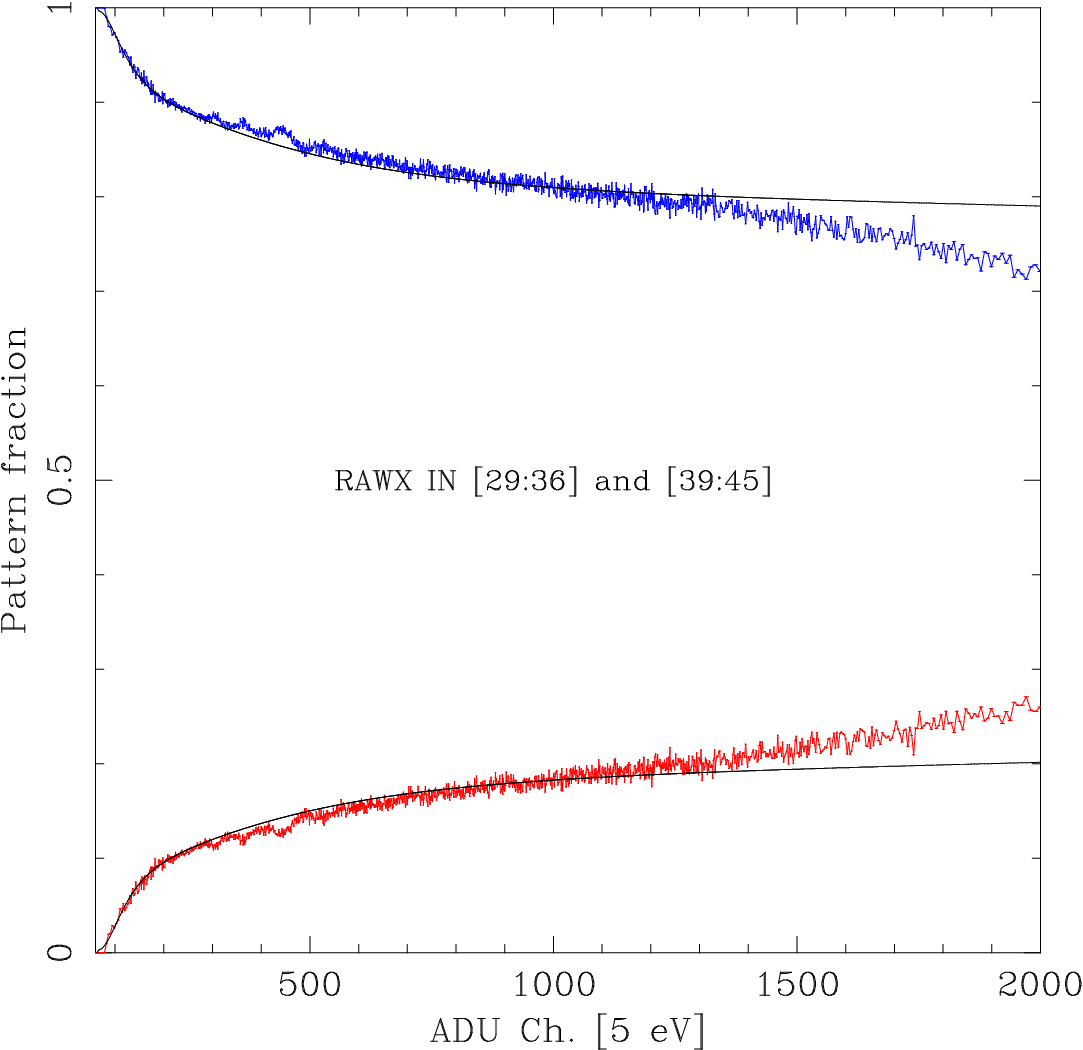}
\caption{Patterns of single (blue color) and double (red color) events 
extracted from the  source region excluding RAWX 37 and 38. 
The patterns are plotted versus
the    60-2000 ADU channel range, corresponding to 0.3-10 keV.}
\label{pileim1}
\end{figure}
finding a photon pile-up percentage of 1.3\%, 2\%, 3.2\%, 5.1\%,
7.4\%, and 9.8\% at 5, 6, 7, 8, 9, and 10 keV, respectively. The
values with the associated errors are reported in Tab. \ref{pileup}.
Since our aim is the study of the Fe-K$\alpha$ region of the spectrum
we investigated how to minimize the photon pile-up effects in the
energy range 5-8 keV. Initially we extracted the source EPIC-pn
spectrum excluding the brightest column (RAWX$=37$) in the CCD finding
that the photon pile-up fraction did not change significantly; we
obtained a photon pile-up percentage of 1.1\%, 1.7\%, 3.1\%, 4.7\%,
6.7\%, and 9.2\% at 5, 6, 7, 8, 9, and 10 keV, respectively. Finally,
excluding the two brightest columns at RAWX$=37$ and RAWX$=38$, we
found a photon pile-up fraction of 0.2\%, 0.6\%, 1.6\%, 1.9\%, 4.8\%,
and 6.5\% at 5, 6, 7, 8, 9, and 10 keV, respectively. In this case the
0.5-12 keV Epic-pn count rate is 920 c/s.  The photon pile-up
percentage values with the associated errors are reported in
Tab. \ref{pileup} and the plot of the fraction patterns of single and
double events are shown in Fig. \ref{pileim1}.  After this
investigation we choose to extract the Epic-pn spectrum excluding the
two brightest columns at RAWX$=37$ and RAWX$=38$ from the source
region to avoid any doubt on the validity of our analysis.

 Since the high available statistics the EPIC-pn spectrum shows evident
calibration issues between 2 and 2.5 keV due to the Au edge near 2.3
keV,  we excluded the EPIC-pn data between 2.1 and 2.5
keV from our analysis. The EPIC-pn spectrum has been rebinned not to
oversample the energy resolution by more than a factor 4 and to have
20 counts per energy channel. The adopted Epic-pn energy range is 0.7-2.1 keV
and 2.5-10 keV.  The spectrum is fitted using XSPEC version 12.3.1

The extracted RGS1 spectrum does not show instrumental issues and we
select data in the 0.6-2 keV energy band for our analysis. The
extracted RGS2 spectrum shows calibration problems in CCD 9 and we
will not use data from this instrument for our analysis.  Since we are
interested in the study of the Fe-K$\alpha$ region, in the following
we concentrate out spectral analysis on the Epic-pn data, after having
checked that the RGS1 spectrum was consistent with the Epic-pn data and
does not show more features.

\begin{figure}[h]
\includegraphics[height=8.cm,angle=0]{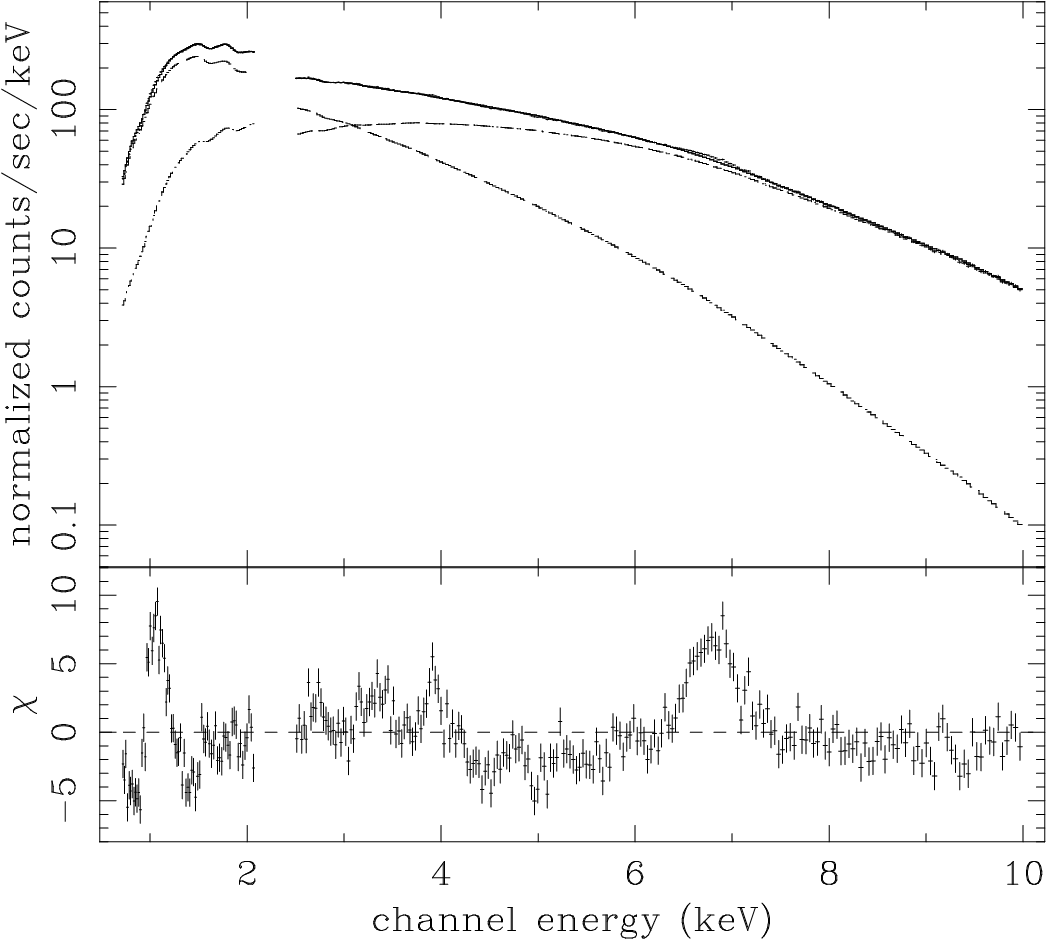} 
\caption[]{Residuals (in units of $\sigma$) in the 0.7-10 keV energy
  range.  The adopted continuum is composed by a blackbody plus a
  multicolored disk blackbody both absorbed by neutral matter.  In the
  residuals clear emission features are evident at 1, 2.7, 3.35, 3.9 and
  in the Fe-K$\alpha$ region of the spectrum. }
\label{fig:res_cont}
\end{figure}

The continuum emission is fitted adopting the same model recently
proposed by \cite{Cackett08_349} for the  Suzaku spectrum of GX
349+2: i.e.  a blackbody component plus a multicolored disk blackbody
(DISKBB in XSPEC) both absorbed by neutral neutral matter (WABS in
XSPEC); because of the narrower energy range of our Epic-pn data (0.7-10
keV) with respect to the Suzaku data we did not add to the model a
power-law component necessary to fit the Suzaku data above 10 keV.  We
obtained a reduced $\chi^2_{red}= 7.73$ with 282 degree of freedom
(d.o.f.).  The residuals, plotted in Fig.  \ref{fig:res_cont}, suggest
the presence of broad emission lines at 1, 2.7, 3.35, 3.9 keV and,
finally a broader emission line between 6 and 7 keV.
Initially, we added five Gaussian components to fit these emission
features. Since the line around 2.7 keV is near the hole in the
Epic-pn data, we fixed the energy of the line at 2.62 keV, imposing
that it is associated to the Ly-$\alpha$ \ion{S}{xvi} transition,
furthermore we fixed its width at 140 eV.  We obtain a
$\chi^2_{red}(d.o.f.)= 1.51(269)$ and a $\chi^2$ improvement of
$\Delta \chi^2 = 1775$. The significance in units of $\sigma$ of the
lines at 1, 2.62, 3.32, 3.9, and 6.7 keV are 14, 7, 15, 9, and 13,
respectively.
    \begin{figure}[h]
\includegraphics[height=7.cm,angle=0]{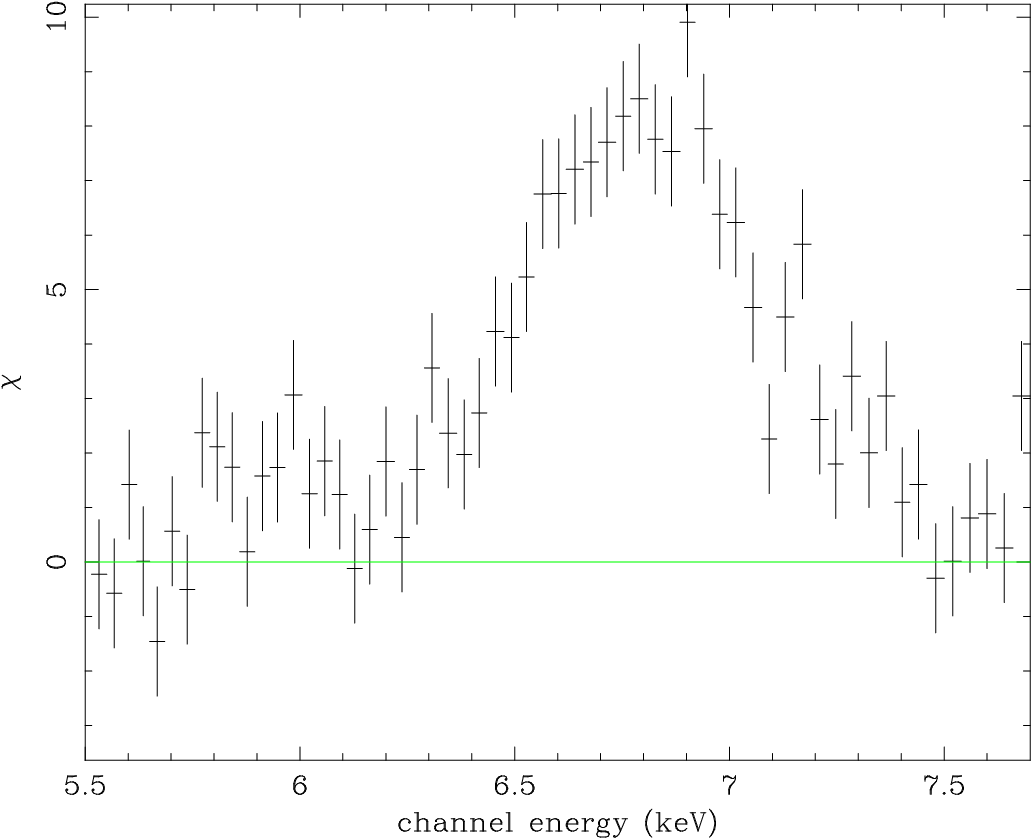} 
\caption[]{Residuals in the 5.5-7.7 keV  energy range. These can be fitted
  adopting a Gaussian profile or, equivalently, a relativistic smeared
  line (diskline in XSPEC). }
\label{fig:res_line}
\end{figure}
The best-fit values are reported in Tabs. \ref{table:model_gauss}
(column 2) and \ref{table:model_gauss_doppia} (column 2).
We identify the broad emission lines centered at
$1.051^{+0.006}_{-0.015}$ keV, $3.33^{+0.03}_{-0.04}$ keV, and
$3.93^{+0.03}_{-0.02} $ keV as a \ion{Fe}{xxii} transition, Ly-$\alpha$
\ion{Ar}{xviii} transition, and resonance transition of
\ion{Ca}{xix}, respectively. The centroid of the broader line between
6 and 7 keV is not identifiable because we find a line centered
at $6.80 \pm 0.02$ keV. We also note that, while
the FWHMs of the broad lines below 6 keV are between 0.2 and 0.4 keV,
the  broader line in the Fe-K$\alpha$ region has a FWHM of
0.6--0.7 keV (the line profile is visible in the residuals shown in
Fig. \ref{fig:res_line}). 

We have also investigated the possibility that the Fe-K$\alpha$
feature is a blending of lines.  We fitted the broad Fe-K$\alpha$
emission feature as a blending of \ion{Fe}{xxv} (6.7 keV) and
\ion{Fe}{xxvi} (6.97 keV) emission lines adopting two Gaussian
profiles and fixing their energies at the expected rest-frame values;
the values of the fit are reported in Tabs. \ref{table:model_gauss}
(column 3) and \ref{table:model_gauss_doppia} (column 3).  Even in
this case the \ion{Fe}{xxv} and \ion{Fe}{xxvi} emission lines appear
to be broad, having widths of 200 eV (FWHM$=$0.5 keV).

Under the hypothesis
that the broadening of the lines is produced by the relativistic
motion of the emitting plasma in the accretion disk close to the
neutron star, we fit the line profiles at 2.62, 3.32, 3.90,
and 6.80 keV with relativistic disk line components (DISKLINE in
XSPEC).  Assuming that the emission features are produced in the same
region, we impose that the inner radius, the outer radius, the
emissivity index, and the inclination angle of the system are the same
for all these components.  Adding these components to the continuum
emission we find a slightly better fit, $\chi^2_{red}(d.o.f.)= 1.47(268)$.
 \begin{table}[h]
  \caption{Broad Gaussian Lines between 6 and 7 keV.}
\label{table:model_gauss_doppia}      
\begin{center}
\begin{tabular}{l c c}          
\hline\hline                        
Parameters  &
Single Fe-K$\alpha$ &
\ion{Fe}{xxv}-\ion{Fe}{xxvi}  \\
  &
Em. Line &
Blending \\

\hline                        
&&\\

E$_{Broad}$ (keV) &
$6.80 \pm 0.02$ &
-- \\

$\sigma_{Broad}$ (eV) &
$280^{+30}_{-40}$  & 
--  \\

I$_{Broad}$ (10$^{-3}$ ph cm$^{-2}$ s$^{-1}$) & 
$4.7^{+0.4}_{-0.6}$   &
--   \\

Equiv. width (eV) & 
$49^{+6}_{-7}$&
--    \\
significance ($\sigma$) &
13 & 
-- \\
&&\\

E$_{Fe \; XXV}$ (keV) &
-- &
6.70 (fixed) \\

$\sigma_{Fe \; XXV}$ (eV) &
--  & 
$220^{+20}_{-40}$  \\

I$_{Fe \; XXV}$ (10$^{-3}$ ph cm$^{-2}$ s$^{-1}$) & 
--   &
$2.6^{+0.7}_{-0.6}$   \\

Equiv. width (eV) & 
--&
$25 \pm 9$    \\
significance ($\sigma$) &
-- &
7 \\
&&\\

E$_{Fe \; XXVI}$ (keV) &
-- &
6.97 (fixed) \\

$\sigma_{Fe \; XXVI}$ (eV) &
--  & 
$280^{+130}_{-70}$  \\

I$_{Fe \; XXVI}$ (10$^{-3}$ ph cm$^{-2}$ s$^{-1}$) & 
--   &
$2.0^{+1.0}_{-0.6}$   \\

Equiv. width (eV) & 
--&
$21 \pm 2$    \\
significance ($\sigma$) &
-- &
5 \\

\hline                                             
\end{tabular}
\end{center}

{\small \sc Note} \footnotesize---    See Tab. \ref{table:model_gauss}. 
\end{table}
   \begin{figure}[h]
\includegraphics[height=6.5cm, angle=0]{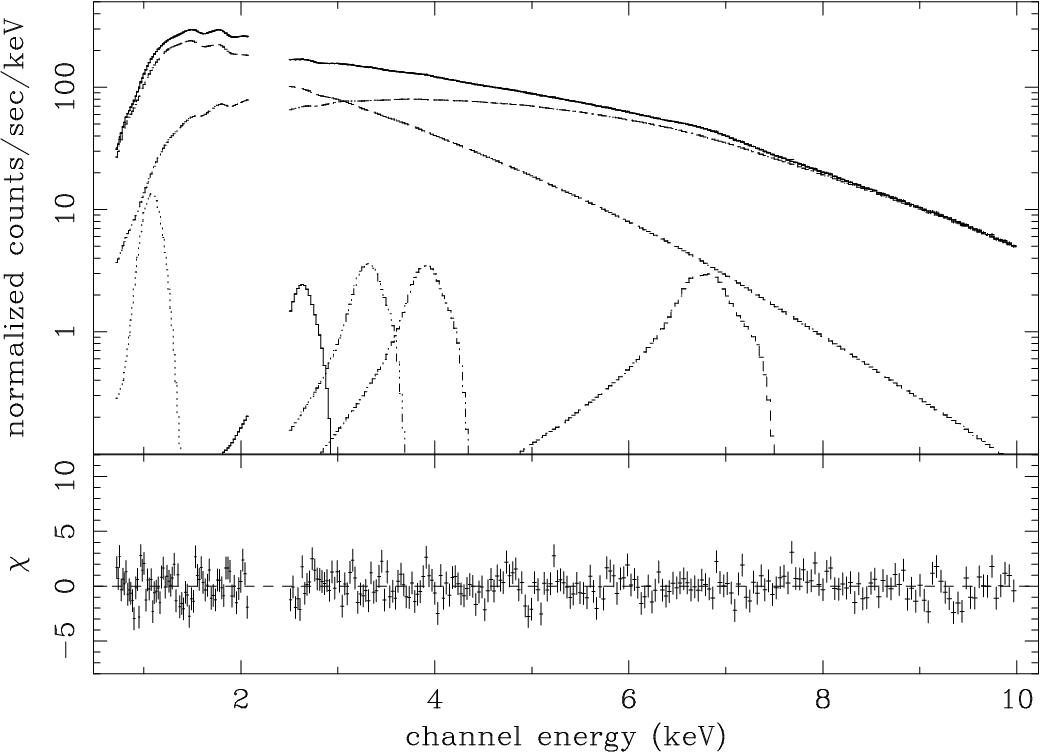}
\caption[]{Data, model, and residuals (in units of $\sigma$) in the 0.7-10  keV energy 
range after adding  the  disk line components discussed in the
  text.}
\label{fig:res_cont_fin}
\end{figure}

\begin{table}[ht]
  \caption{Best-fit values of the parameters obtained fitting the emission features with relativistic smeared lines (diskline in XSPEC). The Fe-K$\alpha$ 
feature is modelled with one diskline component.}
\label{table:model}      
\begin{center}                                      
\begin{tabular}{l c c c }          
\hline\hline                        
Parameters  &
Aver. spectrum&
Spectrum 1+3&
Spectrum 2\\
\hline                        
&
&
&\\

N$_H$  (10$^{22}$ cm$^{-2}$)&  
$0.735 \pm 0.004$&
0.735 (fixed)  &
0.735 (fixed)
\\

&\\

kT$_{BB}$ (keV)         &
$1.797^{+0.016}_{-0.007}$&
$1.780^{+0.023}_{-0.009}$&
$1.818^{+0.013}_{-0.010}$\\

N$_{BB}$ (10$^{-1}$) &
$1.048^{+0.004}_{-0.013}$&
$0.924^{+0.013}_{-0.017}$&
$1.126^{+0.007}_{-0.009}$\\
&\\

kT$_{DISKBB}$ (keV)         &
$1.049^{+0.028}_{-0.014}$&
$1.08^{+0.03}_{-0.02}$&
$1.05 \pm 0.02$\\

N$_{DISKBB}$  &
$207^{+9}_{-20}$&
$213^{+14}_{-18}$&
$190^{+5}_{-12} $\\
&\\

E$_{1 keV}$ (keV) &
$1.052^{+0.006}_{-0.009}$&
 $1.030^{+0.009}_{-0.016}$  &
 $1.062^{+0.004}_{-0.009}$  \\

$\sigma_{1 keV}$ (eV) &
$85^{+13}_{-7}$   &
$120^{+20}_{-8}$  &
$66^{+10}_{-7}$  \\

I$_{1 keV}$  & 
$24^{+4}_{-2}$   &
$42^{+8}_{-4}$   &
$17.0^{+2.1}_{-1.4}$   \\

Equiv. width (eV) & 
$22 \pm 2$  &
$32 \pm 2$  &
$17 \pm 2$  \\

&&\\

$\beta$  emiss. index& 
$-2.1 \pm 0.2               $ &
$-2.10^{+0.12}_{-0.23} $ &
$-2.10^{+0.11}_{-0.16}$ \\

R$_{in}$ ($R_g$) & 
$6.2^{+19.1}_{-0.2}$ &
$6.6^{+12.2}_{-0.6}$ &
$10^{+14}_{-4}$ \\

R$_{out}$ ($R_g$) & 
$1980^{+2900}_{-710}$&
1980 (fixed)&
1980 (fixed)
 \\
$\theta$ (degrees)   &
$41.4^{+1.0}_{-2.1}$  &
41.4 (fixed)&
41.4 (fixed)\\
&\\

E$_{S \; XVI}$ (keV) &
2.62 (fixed)  &
2.62 (fixed)  &
2.62 (fixed)  \\

I$_{S \; XVI}$   &
$2.2^{+0.5}_{-0.9}$  &
$3.4^{+0.9}_{-1.0}$  &
$0.9 \pm 0.6$  \\

Equiv. width (eV) & 
$5 \pm 3$  &
$8^{+4}_{-3} $  &
$2^{+3}_{-2} $  \\

&\\
E$_{Ar \; XVIII}$ (keV) &
$3.29 \pm 0.03$  &
$3.32 \pm 0.02$  &
$3.28^{+0.02}_{-0.04}$  \\

I$_{Ar \; XVIII}$    &
$3.5^{+0.5}_{-0.6}$ & 
$3.8^{+0.9}_{-1.1}$  &
$2.7^{+0.6}_{-0.8} $  \\

Equiv. width (eV) & 
$11 \pm 4$  &
$12 \pm 4 $  &
$9 \pm 4$  \\

&\\

E$_{Ca \; XIX}$ (keV) &
$3.89 \pm 0.03$ &
$3.89^{+0.03}_{-0.04}$ &
$3.90 \pm 0.03$ \\

I$_{Ca \; XIX}$  & 
$3.6^{+0.5}_{-0.7}$   &
$3.5 \pm 0.9$   &
$3.2 \pm 0.7$   \\

Equiv. width (eV) & 
$15 \pm 4$  &
$15 \pm 5$  &
$13 \pm 5$  \\

&\\

E$_{Fe \; XXV}$ (keV) &
$6.76 \pm 0.02$ &
$6.74 \pm 0.04$ &
$6.78 \pm 0.03$ \\

I$_{Fe \; XXV}$  & 
$5.7^{+0.5}_{-0.6}$   &
$5.5 \pm 0.8$   &
$5.8^{+0.7}_{-0.8}$   \\

Equiv. width (eV) & 
$61 \pm 9$  &
$64 \pm 11$  &
$59 \pm 11$  \\

&\\

$\chi_{red}^2(d.o.f.)$ &
$1.47(268)$  &
$1.24(271)$  &
$1.30(271)$  \\

\hline                                             
\end{tabular}
\end{center}                                      

{\small \sc Note} \footnotesize---    Uncertainties are  at the  90\% 
confidence
    level for a single parameter.   
The parameters are defined as in XSPEC. The continuum emission is fitted with 
a {\tt blackbody} plus a multicolored disk blackbody, {\tt diskbb}. 
The four broad emission features are modelled adopting {\tt diskline} 
components.  All components are subject to interstellar absorption. 
The intensity of the lines are in units of 10$^{-3}$ ph cm$^{-2}$ s$^{-1}$.
\end{table}

\begin{table}[ht]
  \caption{Best-fit values of the parameters obtained fitting the emission features with relativistic smeared lines (diskline in XSPEC). The Fe-K$\alpha$ 
feature is modelled with two diskline components.}
\label{table:model_2_diskline}      
\begin{center}                                      
\begin{tabular}{l c c c }          
\hline\hline                        
Parameters  &
Aver. spectrum&
Spectrum 1+3&
Spectrum 2\\
\hline                        
&
&
&\\

N$_H$  (10$^{22}$ cm$^{-2}$)&  
$0.734^{+0.002}_{-0.004}$&
0.734 (fixed)  &
0.734 (fixed)
\\

&\\

kT$_{BB}$ (keV)         &
$1.801^{+0.013}_{-0.009}$&
$1.784^{+0.025}_{-0.013}$&
$1.820^{+0.012}_{-0.011}$\\

N$_{BB}$ (10$^{-1}$) &
$1.043^{+0.007}_{-0.011}$&
$0.920^{+0.013}_{-0.020}$&
$1.123^{+0.008}_{-0.010}$\\
&\\

kT$_{DISKBB}$ (keV)         &
$1.06 \pm 0.02$&
$1.08^{+0.03}_{-0.02}$&
$1.056^{+0.011}_{-0.017} $\\

N$_{DISKBB}$  &
$201^{+5}_{-14}$&
$210^{+14}_{-20}$&
$186^{+10}_{-5} $\\
&\\

E$_{1 keV}$ (keV) &
$1.052^{+0.006}_{-0.005}$&
 $1.032^{+0.009}_{-0.017}$  &
 $1.062^{+0.007}_{-0.009}$  \\

$\sigma_{1 keV}$ (eV) &
$86^{+13}_{-6}$   &
$120^{+20}_{-8}$  &
$66^{+10}_{-8}$  \\

I$_{1 keV}$  & 
$24^{+3}_{-2}$   &
$41^{+7}_{-4}$   &
$17.0^{+2.1}_{-1.4}$   \\

Equiv. width (eV) & 
$22 \pm 2$  &
$31 \pm 3$  &
$17 \pm 2$  \\

&&\\

$\beta$  emiss. index& 
$-2.2 \pm 0.2               $ &
$-2.17^{+0.14}_{-0.19} $ &
$-2.10^{+0.12}_{-0.19}$ \\

R$_{in}$ ($R_g$) & 
$9^{+17}_{-3}$ &
$6^{+14}$ &
$9^{+20}_{-3}$ \\

R$_{out}$ ($R_g$) & 
$3600^{+18000}_{-1900}$&
3600 (fixed)&
3600 (fixed)
 \\
$\theta$ (degrees)   &
$43^{+4}_{-3}$  &
43 (fixed)&
43 (fixed)\\
&\\

E$_{S \; XVI}$ (keV) &
2.62 (fixed)  &
2.62 (fixed)  &
2.62 (fixed)  \\

I$_{S \; XVI}$   &
$2.0 \pm 0.5$  &
$3.4^{+1.0}_{-1.2}$  &
$0.7 \pm 0.6$  \\

Equiv. width (eV) & 
$5 \pm 3$  &
$8 \pm 4 $  &
$2^{+3}_{-2} $  \\

&\\
E$_{Ar \; XVIII}$ (keV) &
$3.30^{+0.04}_{-0.03}$  &
$3.32 \pm 0.05$  &
$3.29 \pm 0.04$  \\

I$_{Ar \; XVIII}$    &
$3.2^{+0.5}_{-0.8}$ & 
$3.7^{+0.9}_{-1.2}$  &
$2.6^{+0.6}_{-0.7} $  \\

Equiv. width (eV) & 
$10^{+4}_{-10}$  &
$12^{+6}_{-5}$  &
$8 ^{+5}_{-3}$  \\

&\\

E$_{Ca \; XIX}$ (keV) &
$3.896 \pm 0.012$ &
$3.89^{+0.03}_{-0.04}$ &
$3.90 \pm 0.03$ \\

I$_{Ca \; XIX}$  & 
$3.4^{+0.5}_{-0.7}$   &
$3.6^{+1.0}_{-1.1}$   &
$3.0 \pm 0.7$   \\

Equiv. width (eV) & 
$14^{+4}_{-8}$  &
$15^{+5}_{-6}$  &
$12^{+3}_{-6}$  \\

&\\

E$_{Fe \; XXV}$ (keV) &
6.70 (fixed) &
6.70 (fixed) &
6.70 (fixed) \\

I$_{Fe \; XXV}$  & 
$4.2^{+0.4}_{-0.9}$   &
$4.9 \pm 1.3$   &
$3.9^{+1.0}_{-0.9}$   \\

Equiv. width (eV) & 
$40 \pm 10$  &
$60^{+20}_{-30}$  &
$40^{+10}_{-20}$  \\
&\\

E$_{Fe \; XXVI}$ (keV) &
6.97 (fixed) &
6.97 (fixed) &
6.97 (fixed) \\

I$_{Fe \; XXVI}$  & 
$1.6^{+0.6}_{-0.3}$   &
$0.9^{+0.9}_{-0.9}$   &
$2.1^{+0.6}_{-0.7}$   \\

Equiv. width (eV) & 
$17^{+9}_{-17}$  &
$10^{+20}_{-10}$  &
$20 \pm 10$  \\

&\\

$\chi_{red}^2(d.o.f.)$ &
$1.49(268)$  &
$1.25(271)$  &
$1.30(271)$  \\

\hline                                             
\end{tabular}
\end{center}                                      

{\small \sc Note} \footnotesize---   See Tab. \ref{table:model}.  
\end{table}

 \begin{figure}[t]
\includegraphics[height=7.cm,angle=0]{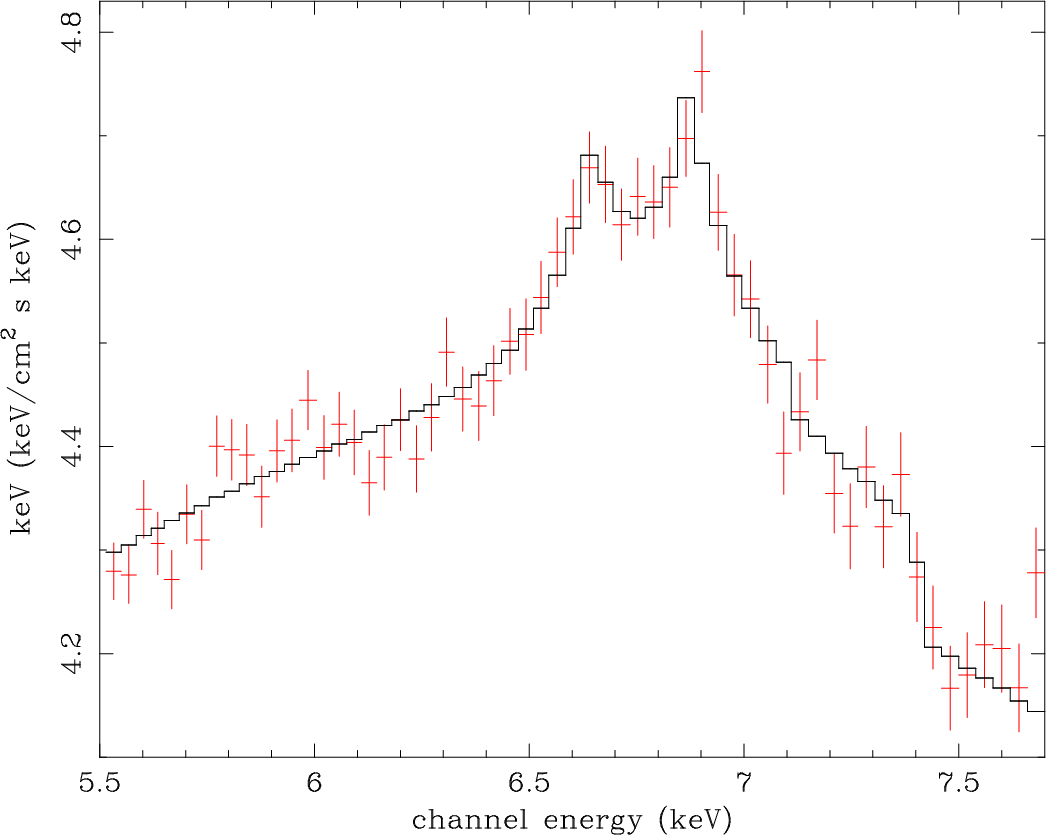}
\caption[]{ Unfolded spectrum in the
  energy  range 5.5-7.7  keV; the  \ion{Fe}{xxv} relativistic  line profile is
  shown.   }
\label{fig:light_eeuf}
\end{figure}
We find that the emissivity index of  the accretion
disk is $-2.1$, the inner radius is less than 25 gravitational radii
($GM/c^2$, hereafter $R_g$), corresponding to 52 km for a NS mass of 1.4
M$_{\odot}$, and the inclination angle of the system is 41$^{\circ}$.
The best-fit values of the parameters are reported in Tab.
\ref{table:model} (column 2); we plot data and residuals in the 0.7-10
keV energy band and the corresponding unfolded spectrum in the
Fe-K$\alpha$ region in Fig.  \ref{fig:res_cont_fin} and Fig.
\ref{fig:light_eeuf}, respectively.

The Fe-K$\alpha$ line has an energy of $6.76 \pm 0.02$ keV and it is
not compatible (at 90\% confidence level) with the rest-frame value of
\ion{Fe}{xxv} transition (6.7 keV).  We therefore tried to fit the
Fe-K$\alpha$ broad feature adopting two diskline components associated
to \ion{Fe}{xxv} and \ion{Fe}{xxvi} transitions. We fixed the energy
values at 6.7 and 6.97 keV, respectively.  The best-fit parameters are
reported in Tab. \ref{table:model_2_diskline} (column 2), in this case
we obtain $\chi^2_{red}(d.o.f.)= 1.48(268)$ and similar parameters of
the emissivity index and inclination angle of the source.  We find an
emissivity index of $-2.2 \pm 0.2$, an inner radius less than 26
$R_g$, corresponding to 52 km for a NS mass of 1.4 M$_{\odot}$, and an
inclination angle of the system of $43^{+4}_{-3}$ deg.

The extrapolated absorbed and unabsorbed flux in the 0.1-100 keV
energy range was $1.2 \times 10^{-8}$ and $1.4 \times 10^{-8}$ erg
cm$^{-2}$ s$^{-1}$, respectively. We obtain an equivalent hydrogen
column density of $0.74 \times 10^{22}$ cm$^{-2}$; the same value was
obtained by \cite{Christian_Swank97} using the Einstein solid-state
spectrometer (SSS; 0.5-4.5 keV), estimating a distance to the
source of $5 \pm 1.5$ kpc (hereafter we will use this distance).
Adopting this value as distance to the source the extrapolated
unabsorbed luminosity in the 0.1-100 keV energy range was $4.3 \times
10^{37}$ erg s$^{-1}$.

\subsection{Spectral analysis of the selected time intervals}
To study possible changes of the broad emission line in the
Fe-K$\alpha$ region of the XMM spectrum of GX 349+2 we extracted the
EPIC-pn spectrum from the interval 2 (Spectrum 2) and from the
intervals 1 and 3 together (Spectrum 1+3). Since the 4-10 keV
lightcurve (Fig. \ref{hardness}, middle panel) indicates a higher
count rate in the time interval 2 than in the time intervals 1 and 3,
we expect a harder spectrum corresponding to the time interval 2. The
lower statistics in the time selected spectra, due to the lower
exposure times, requires that we fix some parameters of the diskline
components. We fixed the outer radius and the inclination angle of the
system to the values obtained fitting the spectrum corresponding to
the whole observation.  Initially we adopted only a diskline component
to fit the Fe-K$\alpha$ broad feature obtaining an energy value of
$6.74 \pm 0.04$ and $6.78 \pm 0.03$ keV corresponding to Spectrum 1+3
and 2, respectively.  We note that while the line energy  is
compatible with a \ion{Fe}{xxv} transition for Spectrum 1+3, it is not
for Spectrum 2, that is when the count rate in the 4-10 keV energy
band increases.  The best-fit values are reported in Tab.
\ref{table:model} (column 3 and 4).

As a consequence of this result we fitted the Fe-K$\alpha$ broad feature
in Spectra 2 and 1+3 adopting two diskline components with energies
fixed at 6.7 and 6.97 keV. Also in this case we fixed the outer radius
and the inclination angle of the system to the values obtained fitting
the average spectrum.  The best-fit
values are reported in Tab.  \ref{table:model_2_diskline} (column 3
and 4).  We find that the main differences between Spectrum 1+3 and
Spectrum 2 is the value of the blackbody normalization that is larger
in Spectrum 2.  Also we note that the flux of the line at 1 keV, the
\ion{S}{xvi} line, the \ion{Ar}{xviii} line, the \ion{Ca}{xix} line,
and \ion{Fe}{xxv} line is lower in Spectrum 2 than Spectrum 1+3 while
the flux of the \ion{Fe}{xxvi} line is larger in Spectrum 2 than
Spectrum 1+3.  In the following section we discuss all these results.

\section{Discussion}

We performed a spectral analysis of a 22.5 ks XMM observation of GX
349+2 in the 0.7-10 keV energy range.  The large flux from the source
caused a telemetry overflow in the EPIC-pn data collection, resulting
an effective exposure time of 10 ks.  We fit the continuum emission
adopting a blackbody plus a multicolored disk blackbody and both these
components are absorbed by neutral matter. Five emission features were
clearly visible in the spectrum.

 Initially we fitted these features using Gaussian profiles and finding
their energies at 1.05, 2.62 (fixed), 3.32, 3.9, and 6.8 keV.  The
corresponding widths and equivalent width were 90, 140 (fixed), 190,
100, and 280 eV, respectively, and 22, 6, 12, 9 and 49 eV, respectively.  We
associated the first four lines to L-shell \ion{Fe}{xxii-xxiii} transition,
Ly-$\alpha$ \ion{S}{xvi}, Ly-$\alpha$ \ion{Ar}{xviii}, and
\ion{Ca}{xix} resonance transition, respectively. The broader emission
feature in the Fe-K$\alpha$ region was not identifiable, hence we
supposed that it is a blending of \ion{Fe}{xxv} and \ion{Fe}{xxvi}
emission lines. Fitting the broad feature adopting two Gaussian
profiles centered at 6.7 and 6.97 keV we obtained that their
corresponding widths were 220 and 280 eV.  We conclude that if the
Fe-k$\alpha$ feature is a blending of \ion{Fe}{xxv} and \ion{Fe}{xxvi}
emission lines, these lines  are intrinsically broad.

BeppoSAX observed three emission lines in the spectrum of GX 349+2
\citep{Iaria04,DiSalvo01}: the first one centered at $1.18 \pm 0.03$ keV,
interpreted as L-shell \ion{Fe}{xxiv} transition, the second at $2.60
\pm 0.06$ keV, interpreted as Ly-$\alpha$ \ion{S}{xvi} transition and,
finally, the third at $6.75 \pm 0.01$ keV, interpreted as
\ion{Fe}{xxv} transition.   In this observation we find that the
centroid of the broad emission line at 1 keV is at a lower energy
suggesting a L-shell transition of less ionized iron
(\ion{Fe}{xxii-xxiii}) although we cannot exclude a more complex blending 
of \ion{Fe}{xxi-xxiv}.
 The widths and equivalent widths of these three
lines are very similar comparing our observation and the two BeppoSAX
observations of GX 349+2 \citep{Iaria04,DiSalvo01} in the non-flaring
state.  \cite{Iaria04} found, during interval 2 in the non-flaring
state, widths of $<80$, $<120$ and $210 \pm 80$ eV, and equivalent
widths of $23 \pm 8$, $7^{+4}_{-1}$, and $52 \pm 11$ eV corresponding
to the \ion{Fe}{xxiv}, \ion{S}{xvi}, and \ion{Fe}{xxv} emission lines.
Furthermore, \cite{DiSalvo01} obtained equivalent widths for the
Fe-K$\alpha$ line of 71 eV and 34 eV, in the non-flaring and flaring
state of the source, respectively.  From these rough comparisons it
seems that the source is in a non-flaring state during the observation
discussed in this work, although we cannot assert this conclusion
without an accurate analysis of the time variabilities.

These emission features can be fitted slightly better adopting smeared
relativistic lines.  The large widths of the lines suggest that these
are produced by reflection of the primary spectral component by the
accretion disk.  Similar broad emission features, discussed as smeared
lines were recently observed in several systems containing a neutron
star, such as Ser X-1 \citep{batta_stromaier07}, SAX J1808.4-3658
\citep{papitto08,cackettsaxj}, 4U 1705-44 \citep{disalvo05, disalvo09}, GX 340+0
\citep{dai08}, 4U 1636-536 \citep{Pandel08}, X 1624-490
\citep{iaria07}, and 4U 1820-30 and GX 349+2 \citep{Cackett08_349}, and
authors of these papers agree to interpret the broadening as due to
relativistic and Doppler effects.  In the following we show the
self-consistency of the disk reflection scenario discussing the
parameters reported in Tab.  \ref{table:model_2_diskline}.

 We infer the inner radius of the accretion disk, $R_{disk}$, using the
normalization value of the multicolored disk blackbody and the
inclination angle of the source, $i=43^{+4}_{-3}$ deg, obtained by the
relativistic smeared line, finding that $R_{disk}$ is $8 \pm 3$ and $8
\pm 2$ km for Spectrum 1+3 and 2, respectively. .  The blackbody
radius, $R_{BB}$, inferred by the best-fit values of the blackbody
normalization is $ 4.2 \pm 1.4$ and $ 4.5 \pm 1.4$ km for Spectrum 1+3
and 2, respectively, too small to be associated to the X-ray emission
from the NS surface.

Taking into account possible modifications of
the multicolored disk blackbody component due to electron scattering
\citep{Shakura_suniaev73,White_stella_parmar88}, the measured color
temperature, $T_{col}$, is related to the effective temperature of the
inner disk, $T_{col}= f T_{eff}$, where $f$ is the spectral hardening
factor, and $R_{eff} = f^2 R_{measured}$.  Adopting $f=1.7$ as
estimated by \cite{shitaka95} for a luminosity around 10\% of the
Eddington limit, close to the 25\% of the Eddington limit estimated for
GX 349+2 in our analysis, the corrected value of $R_{disk}$ is
$24 \pm 7$ km.  
The radius of the accretion disk is compatible with the value of the
inner radius obtained by the relativistic smeared lines, that is, 
assuming a neutron star mass of 1.4 M$_{\odot}$, $<41$ km and 
$19^{+41}_{-6}$ km, for Spectrum 1+3 and 2, respectively. In
the same way, correcting the radius of the blackbody component for the
electron scattering, we infer that $R_{BB} = 13 \pm 4$ km.  This
radius is compatible with the NS radius and suggests that
the blackbody component is produced very close  the NS
surface.  In this scenario we directly observe the emission from
inside the inner radius of the accretion disk, that is close the NS
surface, and the blackbody component is interpreted as the direct
emission from a compact corona and/or a boundary layer (BL) located
between the neutron star and the inner radius of the disk.  We should
expect a Comptonized emission from the compact corona and/or BL, but
if it is optically thick the Comptonized component results saturated
and mimics a blackbody emission in the adopted energy range.

We observe, for the first time, reflection from the inner accretion
disk that involves not only the K-transitions of iron but also ions of
lighter elements as Sulfur, Argon and Calcium which all correspond to
a similar ionization parameter.  We observe emission lines associated
to \ion{S}{xvi}, \ion{Ar}{xviii}, \ion{Ca}{xix}, and \ion{Fe}{xxv},
which correspond to a ionization parameter Log$(\xi) \simeq 3$.  A
similar detection was recently reported by \cite{disalvo09} analysing
XMM data of 4U 1705--44.

Using the relation $\xi = L_x/(n_e r^2)$, where $L_x$ is the
unabsorbed luminosity in the 0.1-100 keV, $n_e$ the electron density
of the plasma, and $r$ the distance from the source of the emitting
plasma we can roughly estimate the electron density $n_e$ adopting as
distance from the source the inner and outer accretion disk radius,
$R_{in}$ and $R_{out}$, obtained from the diskline components.  We
obtain that the electron density $n_e$ decreases from $10^{22}$ to $
10^{17}$ cm$^{-3}$ going from the inner to the outer radius. These
values of electron densities are in agreement with the values reported
by \cite{vrtilek93}, who estimated the electron density of an extended
cloud to be $ 10^{22}$ cm$^{-3}$ near the equatorial plane at a
distance from the neutron star surface lower than $10^8$ cm \citep[see
Fig.  2 in][]{vrtilek93}.
       
 From the analysis of the two selected spectra we find that the
extrapolated (not absorbed) flux in the 0.1--100 keV band of the
multicolored disk blackbody is approximatively constant: $(0.61 \pm
0.12)\times 10^{-8}$ and $(0.49 \pm 0.05)\times 10^{-8}$ erg cm$^{-2}$
s$^{-1}$ for Spectrum 1+3 and 2, respectively. The blackbody component
has a larger flux in Spectrum 2 ($0.94 \pm 0.03 \times 10^{-8}$ erg
cm$^{-2}$ s$^{-1}$) than in Spectrum 1+3 ($0.77 \pm 0.02 \times 10^{-8}$
erg cm$^{-2}$ s$^{-1}$).  This suggests that the intensity change
during the observation is driven by the blackbody component. In our scenario
the blackbody component is associated to the emission from a BL around
the compact object and this emission illuminates the surface of the inner
accretion disk producing the reflection lines observed in the
spectra.  We  note that the line fluxes are larger in Spectrum
1+3 than in Spectrum 2 except for the flux associated to the
\ion{Fe}{xxvi} line which shows an anticorrelated behaviour.

Our scenario may easily explain these results: in fact, if the
emission from the BL increases, the flux illuminating the accretion disk
increases, then the ionization parameter of the reflecting
matter also increases under the hypothesis that the density of the
matter does not change.  This implies that transitions of heavier ions
are more probable, since lighter ions as \ion{S}{xvi},
\ion{Ar}{xviii}, \ion{Ca}{xix}, and \ion{Fe}{xxv} decrease in number
while heavier ions as \ion{Fe}{xxvi} increase in number, giving as
results a larger flux of the \ion{Fe}{xxvi} line and a lower flux of
the other lines.

GX 349+2 is a NS LMXBs belonging to the Z-class sources and its
behaviour is very similar to the prototype of its class Sco X-1.
\cite{steeghs02} analysed the optical spectrum of Sco X-1 and obtained
a firm mass ratio limit of $q \la 0.61$ from the phase-resolved
spectroscopy, where $q$ is the ratio of the companion star mass to the
neutron star mass. If we assume that the optical periodicity of $21.9
\pm 0.4$ hours, observed in GX 349+2 by \cite{wachter96}, is the orbital
period of the system, the mass function of GX 349+2 is $f(M)= 0.032$
M$_{\odot}$ \citep[see][]{wachter96}. We find that the inclination
angle of the system is $43^{+4}_{-3}$ deg. Assuming a neutron star
mass of 1.4 M$_{\odot}$ we find  $q \sim 0.56$, a value very
similar to the upper limit estimated for Sco X-1,
giving a mass of the companion star of $M_c = 0.78^{+0.07}_{-0.06}$
M$_{\odot}$, typical for a NS LMXB of the Z class ($q \le 0.61$ for
Sco X-1, \citealp{steeghs02}; $q = 0.34 \pm 0.04$ for Cyg X-2,
\citealp{casares97}).

Furthermore Sco X-1 shows twin compact radio lobes forming an angle, 
with respect the line of sight, of $44 \pm 6$ deg
 \citep[error at 1 $\sigma$, see][]{fomalont01}. If the jet from the source
producing the radio lobes is almost perpendicular to the equatorial
plane of the system then the angle of $44 \pm 6$ deg corresponds to
the inclination angle of the system, similar to the inclination angle
of $43^{+4}_{-3}$ deg obtained  for GX 349+2  in this work. Our results
indicate that Sco X-1 and GX 349+2 are very similar sources, as already
suggested by other authors \citep[see e.g.][]{KuulVdk}.

Although our scenario looks reasonable, our values of the diskline
parameters are not compatible with the recent ones reported by
\cite{Cackett08_349}.  \citeauthor{Cackett08_349} analysing Suzaku
data of GX 349+2 fitted the broad emission feature in the Fe-K$\alpha$
region adopting a smeared relativistic line with the outer radius
fixed at 1000 $R_g$.  They found a line energy associated to
\ion{Fe}{xxvi}, an inner radius of $16.5 \pm 0.8$ km, an emissivity
index of $-4.1 \pm 0.3$ and, finally, an inclination angle of the
system of $23^{\circ} \pm 1^{\circ}$.  The differences in line energy,
emissivity index, and inner radius of the reflecting skin could be
explained supposing that GX 349+2 was observed in a different spectral
state.  We find that the Fe-K${\alpha}$ broad emission feature is
associated to \ion{Fe}{xxv} line or to a blending of \ion{Fe}{xxv} and
\ion{Fe}{xxvi} in which the \ion{Fe}{xxv} line is more intense.  Our
measurement is similar to the previous BeppoSAX observations of GX
349+2 \citep{DiSalvo01,Iaria04} and to the two recent Chandra
observations of this source \citep{cackett09}. In all these
observations GX 349+2 was in NB/FB and/or in the bottom part of the
FB. The only exception is the Suzaku observation of GX 349+2
\citep{Cackett08_349} where the broad emission line seems associated
to \ion{Fe}{xxvi} suggesting that the source was in a different state;
in fact, as discussed by \citeauthor{Cackett08_349}, it is possible
that GX 349+2 was in NB during that observation.

However, the different inclination angles of the system
obtained in this work and by \cite{Cackett08_349} cannot be
easily explained since the inclination angle of the system should
not depend on the state of the source and should not  change. Using
the inclination angle obtained by \citeauthor{Cackett08_349} we obtain
a mass ratio $q \sim 1.25$ and a companion star mass of 
$1.74^{+0.12}_{-0.11}$ M$_{\odot}$ for a neutron star mass of 1.4
M$_{\odot}$, that is quite unusual for a NS LMXBs. 

 
Higher statistics spectroscopic studies are needed to put tighter constraints 
on the system parameters of  this source, since these can give 
important informations on the whole binary system and on the physics 
of the accretion flow close to the compact object.

\section{Conclusions}

We analysed a XMM observation of GX 349+2 finding the presence of
prominent emission features associated to \ion{S}{xvi},
\ion{Ar}{xviii}, \ion{Ca}{xix},  \ion{Fe}{xxv} and, possibly,  \ion{Fe}{xxvi}.  
The emission
features can be fitted with relativistic smeared lines ({\tt
  diskline}).  The continuum emission was fitted adopting a
multicolored disk blackbody plus a blackbody component.  We
investigated the scenario in which the broad emission lines are formed 
 by reflection from a ionized skin in the inner region of the accretion disk
illuminated by the emission of a compact corona (or a boundary layer) 
surrounding the
neutron star that we fitted with a blackbody component with a
temperature of 1.8 keV. We find a self-consistent explanation of our
results.  The inner radius of the  disk blackbody is $24 \pm 7$ km
while the broad lines give an inner disk radius  less than  40 km
from the neutron star center.  We discuss
that the electron density of the reflecting plasma is
between $10^{17}$ and $10^{22}$ cm$^{-3}$ and  that the
inclination angle of the source is around 43$^{\circ}$. Finally, we
infer that, for an inclination angle of 43$^{\circ}$, the mass of
the companion star is 0.78 M$_{\odot}$.

\begin{acknowledgements} 
We are very grateful to the referee for his/her suggestions and comments.
This paper is published despite the effects of the Italian law 133/08.
This law drastically reduces public founds to public Italian universities, 
which is particularly dangerous for scientific free research, and it will 
prevent young researchers from getting a position, either temporary or 
tenured, in Italy. The authors are protesting against this law to obtain 
its cancellation.
\end{acknowledgements} 
\bibliographystyle{aa} 
\bibliography{citations}
\end{document}